\documentclass[12pt]{emulateapj}

\newcommand{\xmm}{\emph{XMM-Newton}}
\newcommand{\chisq}{$\chi^2$}
\newcommand{\cm}[1]{~cm$^{#1}$}
\newcommand{\cms}{~cm$^{-3}$\,s}

\newcommand{\e}[1]{10$^{#1}$}
\newcommand{\ee}[1]{$\times$10$^{#1}$}
\newcommand{\ergs}{~ergs\,cm$^{-2}$\,s$^{-1}$}
\newcommand{\hi}{H\,{\sc i}}

\newcommand{\msun}{M$_{\odot}$}
\newcommand{\nh}{N$_{\rm H}$}
\newcommand{\oh}{OH(1720~MHz)}

\shorttitle{Thermal X-ray emission from the SNR IC443}
\shortauthors{E. Troja et al.}

\begin{document}
\title{
{\it{XMM-Newton}} observations of the supernova remnant IC443:
I. soft X-ray emission from shocked interstellar medium}

\author{E. Troja\altaffilmark{1}, F. Bocchino\altaffilmark{2},
        and F. Reale\altaffilmark{1,2}}

\altaffiltext{1}{Dipartimento di Scienze Fisiche ed Astronomiche,
Sezione di Astronomia, Universit\`a di Palermo, Piazza del
Parlamento 1, 90134 Palermo, Italy}
\altaffiltext{2}{INAF-Osservatorio Astronomico di Palermo,
Piazza del Parlamento 1, 90134 Palermo, Italy}

\begin{abstract}

The shocked interstellar medium around IC443 produces strong X-ray
emission in the soft energy band (E $\lesssim$ 1.5 keV). We present an
analysis of such emission as observed with the EPIC MOS cameras on
board the \xmm\ observatory, with the purpose to find clear 
signatures of the interactions with the interstellar medium (ISM)
in the X-ray band, which may complement results obtained in other wavelenghts.\\ 
We found that the giant molecular cloud mapped in CO emission is located 
in the foreground and gives an evident signature in the absorption of X-rays. 
This cloud may have a torus shape and the part of torus interacting with 
the IC443 shock gives rise to 2MASS-K emission in the southeast. 
The measured density of emitting X-ray shocked plasma increases 
toward the northeastern limb, where the remnant is interacting with an atomic cloud. 
We found an excellent correlation  between emission in the 0.3--0.5~keV band 
and bright optical/radio filament on large spatial scales. 
The partial shell structure seen in this band therefore traces the 
encounter with the atomic cloud.

\end{abstract}

\keywords{X-rays: ISM---ISM: supernova remnants---ISM: individual
object: IC 443}

\section{Introduction}\label{sec:intro}

SNR IC443 (G189.1+3.0) is located in the Gem~OB1 association at a
distance of 1.5~kpc \citep{welsh03}. Optical and radio morphology is
shell-like, it has an angular diameter of 50' and appears to consist
of two connected subshells with different radii \citep{braun86}. In
the X-ray band IC 443 has been previously observed with {\it
Einstein} and the HEAO 1 A2 experiment \citep{petre88}, Ginga
\citep{wang92}, ROSAT \citep{asaoka94}, ASCA \citep{keohane97,
kawasaki02}, BeppoSAX \citep{bocchino00} and RXTE \citep{sturner04}.

IC443 shows features similar to the class of mixed morphology
supernova remnant \citep{mixed98}. The X-ray morphology is centrally
peaked and there is no evidence of a limb-brightened X-ray shell.
The X-ray emission is primarily thermal and it has been described
with a two component ionization equilibrium model of temperatures
$\sim$0.2--0.3~keV (hereafter cold component) and $\sim$1.0~keV
(hereafter hot component, \citealt{petre88,asaoka94}).
On the contrary, the work of \citet{kawasaki02} suggests
a non-equilibrium condition of the X-ray emitting plasma,
reporting a low ionization timescale
for the cold component and a overionization state for the hot one.

Recently {\it Chandra} and \xmm\ observations identified
a plerion nebula in the southern part
of the remnant \citep{olbert01,bocchino01}, but its association with
IC443 is still under debate. Source position seems not compatible at
the 90\% confidence limit with the EGRET source 3EG J0617+2238
\citep{esposito96}, as argued by \citet{bocchino01}.

The environment in which the remnant is evolving seems to be quite
complex. In the south the blast wave has been decelerated by the
encounter with a dense ($n_0$$\sim$10$^4$ cm$^{-3}$) and clumpy
molecular ring \citep{burton88, dickman92}. A rich spectrum of molecular lines emission
(\citealt{vandish93,rho01} and references therein) together
with \oh\ maser emission \citep{claussen97, hoffman03} have been detected,
confirming the interaction with high density gas.
In the northeast the shock front has encountered a less dense
(10$<$$n_0$$<$10$^3$~cm$^{-3}$) \hi\ cloud \citep{denoyer78,rho01}.\\
Emission from quiescent molecular gas has been observed toward IC443
direction and it is likely due to a giant molecular cloud in front of
the remnant \citep{cornett77}. A superposition between IC443 and
another separated SNR (G189.6+3.3) has also been suggested \citep{asaoka94}.\\
Multiwavelenght observations have shown the presence of sharp density
gradients and different clouds geometries in the surroundings of IC443.
Establishing the exact position of ISM structures respect to the IC443
is of crucial importance to understand SNR dynamics and evolution.

In this paper, we focus on the interaction between the SNR shock
and the dense ISM environment. For this reason, we present a detailed analysis of the softer X-ray
thermal emission of the SNR IC443 based on \xmm\ observatory data.
We leave the study of the hot thermal component, which is more prominent 
in the inner regions of the remnant, to a successive paper, now in preparation.
The aim of this work is to take advantage of \xmm\ characteristics
to derive a better description of the SNR evolutionary scenario
and to directly measure physical conditions of the X-ray
emitting plasma both in the bright regions and in the yet unexplored
faint western regions.

In Section~\ref{sec:data}, we present our observations and the data
reduction methods we used. Section~\ref{sec:bgrnd} provides a
description of our background subtraction method. In
Section~\ref{sec:results}, we present the results obtained from the
\xmm\ data analysis. Section~\ref{sec:discuss} contains a discussion
on the physical properties of the cold X-ray emitting plasma
(Sect.~\ref{sec:global}), on the interaction between the remnant and
the surrounding clouds (Sect.~\ref{sec:gradient}) and on the
absorption from a foreground dense molecular cloud
(Sect.~\ref{sec:absorption}). In Section~\ref{sec:end}, we summarize
our results and conclusions.

\section{Data analysis}\label{sec:data}

\subsection{Observations}

IC443 was observed on 2000 September as part of the Cal/PV phase of
the \xmm\ Observatory, using EPIC and RGS instruments. The
observation was performed in four distinct pointings to allow a
maximum coverage of the source.
These are the same observations analyzed by \citet{bocchino03}.

Here, we focus only on the data collected by the EPIC-MOS cameras
\citep{turner01}. During
each observation, the MOS cameras were operated in Full Frame Mode,
providing a temporal resolution of 2.6 s, and the medium filter was
used to limit the number of low-energy photons. The Science Analysis
System software (SAS, version 6.0) was used for data reduction.
Spectral analysis was performed with \textsc{xspec}~v.11.1 \citep{arnaud96}.

We generated calibrated events files with the SAS task \emph{emproc} and then
we further screened the data,
selecting only events with PATTERN $\leq 12$ and an optimal value of FLAG
, as suggested by the current status of calibration\footnote{http://xmm.esac.esa.int/external/xmm\_sw\_cal/calib/}.
We discarded events with energy below 0.3 keV from our future analysis 
because of data calibration uncertainties.


\begin{deluxetable}{ccccc}
\tabletypesize{\scriptsize} \tablewidth{0pt} \tablecaption{General
informations about MOS data
(MOS1+MOS2).\label{tab:obsid}} 
\tablehead{ \colhead{OBS\_ID}  & \colhead{Date} & \colhead{$\alpha_{J2000}$} & \colhead{$\delta_{J2000}$} & \colhead{T$_{exp}$\tablenotemark{a}}\\
            \colhead{}         & \colhead{UT}     & \colhead{(hh:mm:ss)}       & \colhead{(dd:mm:ss)}       & \colhead{(ksec)}
} \startdata

 0114100101     & 26/09/2000
                & 06:17:28
                & +22:41:44
                & 20/45 \\
 0114100201 & 25/09/2000
                & 06:16:15
                & +22:41:60
                & 10/11 \\
 0114100501             & 25/09/2000
                & 06:16:15
                & +22:41:60
                & 40/50 \\
 0114100601 & 27/09/2000
                & 06:17:28
                & +22:25:44
                & 11/12 \\
 0114100301         & 27/09/2000
                & 06:17:28
                & +22:25:44
                & 40/50 \\
 0114100401     & 28/09/2000
                & 06:16:15
                & +22:18:00
                &  50/60 \\
\enddata
\tablenotetext{a}{Screened/unscreened exposure time}
\end{deluxetable}



\begin{figure}
\epsscale{1.3}
\plotone{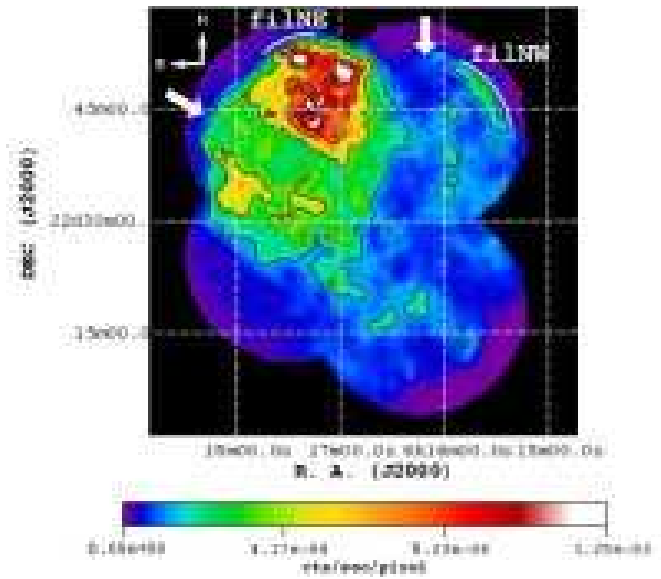}
\caption{X-ray count rate image of
IC443 in the 0.3--1.4 keV energy band. The image was adaptively
smoothed using $\sigma_{min}$=5''and $\sigma_{max}$=20''; the bin
size is 5''. Contour levels correspond to 1.4, 4, 7, 10 10$^{-4}$
cts s$^{-1}$ bin$^{-1}$. Arrows indicate regions of low brightness
surface; the partial shell structures are labeled as filNE and
filNW.}
\label{fig:xima}
\end{figure}


\begin{figure*}
\epsscale{1.25}
\plotone{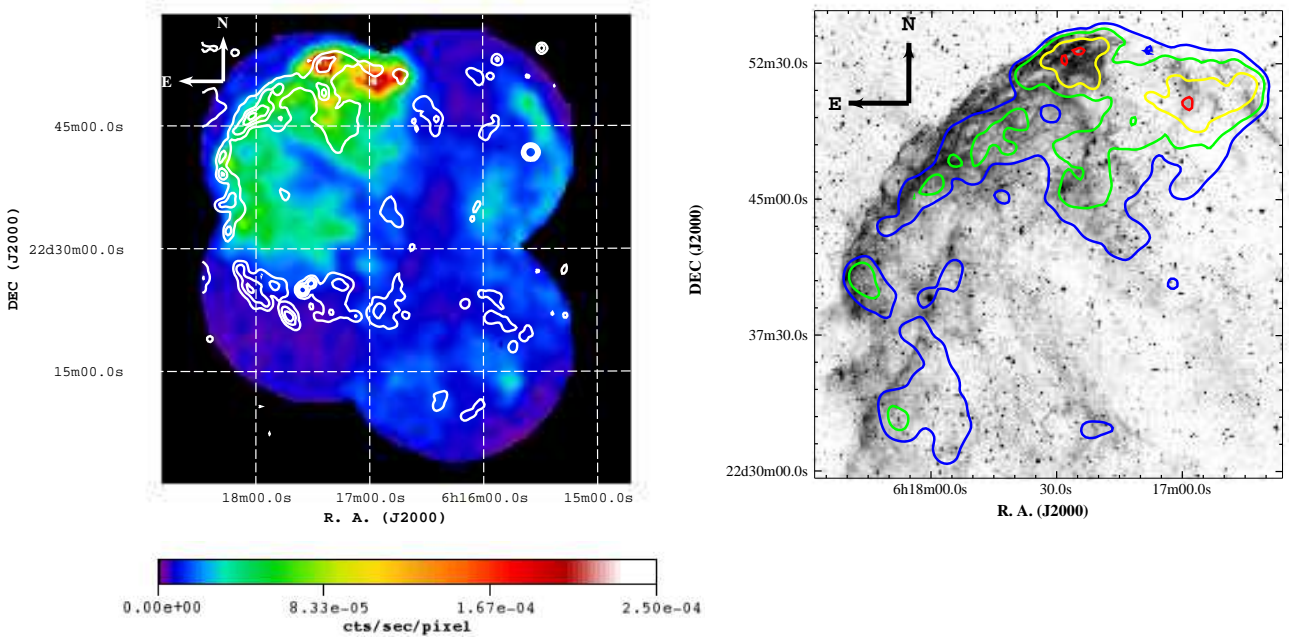}
\caption{\textit{Left panel:} X-ray
count rate image of IC443 in the 0.3--0.5 keV energy band. The image
was adaptively smoothed using $\sigma_{min}$=10'' and
$\sigma_{max}$=40''; the bin size is 10''. We overlaid VLA NVSS
contours at 1.4 GHz, corresponding to 5, 20, 45, 80 mJy
beam$^{-1}$. \textit{Right panel:} Optical DSS image of the northeastern
shell of the remnant overlaid with contours of X-ray emission in the
0.3--0.5 keV energy band. Contour levels correspond to 1.7 (blue), 2.6
(green), 4.6 (yellow), 8.6 (red) \e{-3} cts s$^{-1}$ arcmin$^{-2}$.}
\label{fig:xima2}
\end{figure*}


\subsection{Background subtraction}\label{sec:bgrnd}

The EPIC MOS background components are mainly electronic noise, fluorescence emission lines
(instrumental background, not vignetted), X-ray galactic emission
(astrophysical background, vignetted) and soft magnetospheric
protons focused by the optics \citep{lumb02,marty03}

We removed high proton background time intervals by
extracting light curves at energies $> 8$ keV, where the EPIC
cameras have a low quantum efficiency, with a bin time of 30~s .
We rejected data intervals with more than 15 counts.

To check for a residual proton component we used a simple but useful
diagnostic, developed by \citet{demol04}. We calculated
for each screened data set the ratio R between the surface
brightness in FOV ($\Sigma_{IN}$) and the surface brightness out FOV
($\Sigma_{OUT}$) in the energy range 8-12 keV. Following the
definition given in \citet{demol04}, the maximum value of
R, above which there is a significant soft proton contamination, is
R$_{max}$=1.3. If 1.05$<$R$<$1.3 the residual proton level is low,
if R$<$1.05 it can be considered negligible.
We verified that our cleaned data sets have all a low or
negligible residual proton component.
After the screening process, the total exposure time was reduced by 25\% (from 230 to 170
ks).

Table~\ref{tab:obsid} summarizes the main information about the
observation: the date, the pointing's center coordinates, the
exposure time before and after flare screening.

Instrumental background is the dominant component of the events out of FOV
and it is also highly spatial variable. A typical solution
is to estimate it using merged data sets acquired with the filter
wheel in the CLOSED position \citep{lumb02}.
We used the closed data sets of P.~Marty, distributed by
the School of Physics and Astronomy of the University of Birmingham\footnote{http://www.sr.bham.ac.uk/xmm3/scripts.html}.

Since IC443 is an extended source, that fills the entire field of
view, it is not possible to estimate the local astrophysical
background from the same data. Instead of a local background, we
used a set of observations of blank sky regions, produced from the
Galactic Plane Survey observations (GPS,~P.I.~Parmar), pointed toward a galactic
longitude of 319$^\circ$ and a galactic latitude between 0$^\circ$ and 3$^\circ$.
This data set represents a better estimate of the cosmic background
at low galactic latitude than the standard blank sky field at high $|b|$.\\
After the same screening process described in Sect.~\ref{sec:data}
and the removal of point sources, data sets were scaled by the
relative exposure time and then reprojected onto the sky attitude of
IC443 pointings with the script \emph{skycast}.

From a comparison between the spectra extracted
from the whole FOV of the CLOSED, GPS and IC443 data,
we noted that in the energy range 0.3-5.0 keV the source
signal is the brightest, while above 5.0 keV the instrumental
noise component is the dominant one.\\
We decided to consider in our future analysis only events in the
band 0.3--5.0~keV and to use GPS data as background correction.
We verified that in this band our results are not sensitive
to the background subtraction (either GPS or high galactic latitude
blank fields).

\section{Results}\label{sec:results}
\subsection{X-ray Images}\label{sec:xima}

We defined two energy bands, in which images has been extracted, so that
in each band the contribution of a single component is the dominant
one and, at the same time, the statistic data quality is good. In
the chosen range 0.3--1.4 keV (soft band), the cold component flux
(calculated assuming the parameters of the two components fit 
reported by \citealt{petre88}; \citealt{asaoka94}) is the 64\% of the total; 
in the range 1.4--5.0 keV (hard band), the hot component flux
represents the 90\% of the total.\\
A third image, generated in the 0.3--0.5 keV band, traces the
spatial distribution of the coldest X-ray emitting plasma and it is
useful to trace the boundary between the SNR shock and the
surrounding molecular clouds.

All the X-ray emission images presented here are background
subtracted, exposure and vignetting corrected. To correct for the
exposure and the vignetting, we divided the mosaiced
background-subtracted count image by its mosaiced exposure map. The
resultant image was adaptively smoothed with the SAS task
\emph{asmooth}.

Figure~\ref{fig:xima} shows the 4 pointings mosaiced image of IC443
in the 0.3--1.4 keV band. The brightest features are concentrated in
the northeastern (NE) quadrant and a clear shell is not present. We
mark a region near the rim of the remnant (filNE in
Figure~\ref{fig:xima}): the high surface brightness gradient suggests
that we are looking the shock front nearly edge-on. The large
effective area and the good spatial resolution of \xmm\ allow us to
have a better view of the western part of the remnant and to resolve
its structure for the first time. Western X-ray emission appears
about 10 times fainter and more diffuse than the eastern one. We can
identify a bright filament (filNW in
Fig.~\ref{fig:xima}), that suggests the existence of a
limb-brightened partial shell.\\
The arrows in Fig.~\ref{fig:xima} indicate two dark lanes, that
cross the remnant: the first, from north to south, is probably due
to the presence of a molecular cloud between the observer and IC443
\citep{petre88}; the second, from northeast to southwest, was explained by
\citet{asaoka94} as an additional absorption effect by
the SNR G189.6+3.3.

On the other hand, the morphology of the very soft emission
(0.3--0.5 keV, see Fig.~\ref{fig:xima2}) is not centrally filled,
the X-ray surface brightness peaks are near
the northeastern edge of the remnant, resembling strikingly IC443
radio and optical emission. For comparison, we show in
Fig.~\ref{fig:xima2} (\textit{left panel}) the X-ray image
of the remnant in the 0.3--0.5 keV band, overlaid with
contours of radio emission levels at 1.4 GHz,
observed with the VLA NVSS survey \citep{condon98}.
We also present the superposition between X-ray contours and the optical
image, extracted from the DSS red survey \citep{lasker90}, of the northeastern quadrant of IC443,
where it is encountering a \hi\ cloud \citep{denoyer78}
(Fig.~\ref{fig:xima2}, \textit{right panel}).\\
In the very soft X-ray band it is possible to resolve a partial
shell structure in the NE, similar to that we see in the radio and
optical images. \citet{kawasaki02} had already claimed the presence
of a shell-like structure on the basis of ASCA GIS softness ratio
map in the energy bands 0.7--1.5 and 1.5--10~keV. However our
results represent a direct detection of a very soft shell, which is
likely different from the structure reported by \citet{kawasaki02}.
In the light of our result on molecular cloud obscuration of the
IC443 X-ray emission (see par.~\ref{sec:absorption} below), we argue
that the structure detected by \citet{kawasaki02} in their softness
ratio map is due to the absorption of the foreground molecular cloud
of \citet{cornett77} instead of a real shell morphology.

In the northeastern quadrant, radio, optical and very soft X-ray emission 
correlate well: X-ray brightest features are confined within the radio-optical
shell and the position of maximum X-ray emission contour levels,
just behind bright optical filaments, is in agreement with a shock
front that has just encountered a density gradient of the ISM with
higher density located in the outer regions.
This correlation was also noted by \citet{petre88}, which however
focused only on the northern edge of the brightest soft X-ray region. 
Our results instead shows that this correlation involves the very soft X-ray
shell and the optical-radio emission on a much greater extent.\\
The structures seen in the 0.3-0.5 keV map are of course affected by absorption,
however we will see in sect.\ref{sec:median} that the shape of the very soft shell 
is not correlated with the absorption map in the east of the SNR. 
In the light also of the good correlation between the very soft shell and the
radio emission in the NE of IC443, we argue that a real shell structure exists.

In the southeastern quadrant, where the remnant is interacting with a
dense molecular cloud, the spatial gap between X-ray emission and
radio contours, as shown in Fig.~\ref{fig:xima2}, suggests
that the shock front has been penetrated deeply inside the ring of
molecular gas and it has been strongly decelerated, emitting mainly
in the infrared band \citep{burton88, richter95, rho01}.


\begin{figure}
\epsscale{1.25}
\plotone{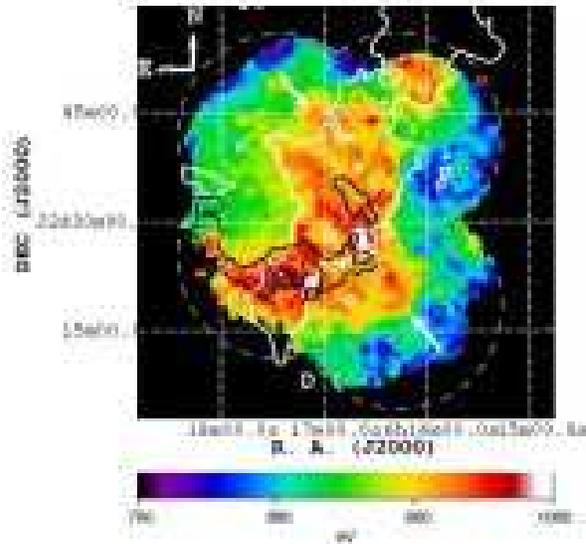} \caption{Median energy map in the soft band
(0.3--1.4 keV), overlaid with quiescent CO (white) and shocked H$_2$
(black) emission contours. The arrows indicate two high median energy regions
without a counterpart in CO or H$_2$ maps.
We masked some regions inside the observed field of view, 
traced by the dashed line, as they are too noisy (S/N$<$3). 
Image bin size is 20'' and smoothing width is $\sigma$=20''. 
The velocity-integrated (-6 to 0 km s$^{-1}$) J=1--0 CO emission 
has 50'' resolution. The velocity-integrated $v$=1--0 S(1) H$_2$ 
line emission at 2.122 $\mu$m has the spatial resolution of 40''
\citep{burton88}.}
\label{fig:q50}
\end{figure}


\begin{figure}
\epsscale{1.2}
\plotone{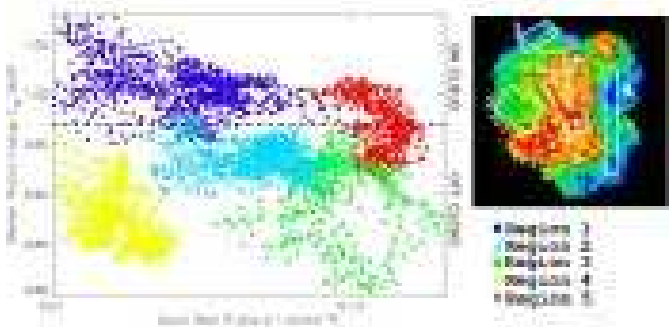} \caption{Median/rate plot for the
soft band (0.3--1.4 keV): on the Y-axis is reported the median
photon energy, on the X- axis the corresponding count rate. 
The dashed line traces the median energy threshold  at 0.97 keV
between ``on cloud" and ``off cloud" regions.
Points come from five regions, superimposed on the median energy map in the
top right corner. 
The black contour in the map corresponds to the value of E$_{50}$=0.97~keV.}
\label{fig:midrate}
\end{figure}


\subsection{Median Energy Map}\label{sec:median}

The complicated morphology of IC443 and the previous literature results
suggest that the plasma has a distribution of
physical conditions and so it is crucial to perform our
spectral analysis selecting regions as homogeneous as possible. The
median photon energy E$_{50}$ is a robust spectral indicator, as shown in
\citet{hong04}, that allows us to map spectral variations inside
the field of view.

For a plasma in ionization equilibrium and with solar abundances
(model MEKAL in XSPEC) we verified that in the
soft band (0.3--1.4 keV) E$_{50}$ is mainly influenced by absorbing column
variations, especially at large values of the
absorption (\nh$\geq$0.5\ee{22}\cm{-2}). The choice of the band
assures that the median energy is not heavily affected by the parameters
of the X-ray hot component.

Fig.~\ref{fig:q50} shows the median energy map of the emitting
plasma in the 0.3--1.4 keV band. Molecular line data toward IC443
are plotted as contours over the image: quiescent CO in white
(Lee~J.~J., private communication) and shocked H$_2$ in black \citep{burton88}.\\
To build the map we used a pixel size of 20'',
in order to collect at least 10 photons per pixel, and then,
for each pixel we calculated the median photon energy.
The resulting image was smoothed using a bidimensional
Gaussian distribution with $\sigma$=20''. No correction for the
X-ray background was made, we simply masked faint regions, where the
signal to noise ratio was less than 3.\\
The image presented in Fig.~\ref{fig:q50} allows us to compare
spectral properties of the cold X-ray component with the emission of
shocked and quiescent molecular gas with an unprecedented spatial
resolution (28'', corresponding at $\sim$0.2~pc at a distance of
1.5~kpc). \\
Previous studies about IC443
\citep{petre88,asaoka94} tried to link spectral hardness ratio variations
across the remnant to an absorption effect, but the low images
resolution did not permit to study the correlation with the molecular cloud 
material on small angular scales. Moreover, previous hardness ratio maps are 
sensitive both to absorption effects and to the parameters of the hot component, 
whereas our median energy map is more strongly dependent on absorption.\\
The spatial variations of the spectral properties reported in our map (Fig.\ref{fig:q50})
confirm the distribution of the hardness ratio in \citet{asaoka94}, but shows more
details.
In particular, we find a strict correlation, also on small angular scales, 
between molecular emission contours and high median energy regions, 
as expected in the case of an increasing column density. 
They match very well everywhere, except two high median regions, indicated by the arrows
in Fig.~\ref{fig:q50}, that seems not to have a counterpart in CO and H$_2$ maps.

As already shown in the study of the Vela supernova remnant
of \citet{miceli05}, the relation between the median photon energy
E$_{50}$ and the corresponding count rate R in each pixel is able to
give a global view of different thermal structures and spectral
variations inside the field of view, preliminary to a complete
spectral analysis. As shown in Fig.~\ref{fig:midrate}, we selected
five significant regions and made the corresponding E$_{50}$ vs. R
plot to get further information about the physical parameters of the
plasma. We applied the Maritz-Jarrett method to estimate the errors on the median
energy, as described by \citet{hong04}, and we assumed a
poissonian error for the count rate values. In order to avoid excessive
noise, especially in faint areas,
points with $\sigma(E_{50})\geq$5\% or
$\sigma(R)\geq$15\% were not reported in the plot.

It is worth to note that all the points belonging to region 1 (blue
circles), which corresponds to a CO emission area, have the highest
median values (E$_{50}\geq$1.0 keV), while points from other regions
lie below this energy threshold. In particular, we can
distinguish two distinct branches at 1.0 keV (region 1) and at 0.9
keV (region~2 and a great part of region 3). These features could be
interpreted as an absorption effect: on average plasma physical
conditions appear to be the same in different regions, the only
difference being an increase of the column density, that shift
points from region 1 towards higher median and lower rate values.
We may define the boundary of off-cloud and on-cloud regions with the
value E$_{50}$=0.97~keV (dashed line in Fig.~\ref{fig:midrate}, {\it left panel}),
whose contour is also reported in the right panel of Fig.~\ref{fig:midrate}.
Data from the off-cloud region 4 (yellow diamonds) are those affected by the
greatest uncertainties. They show similar E$_{50}$ values of region 3 (green squares),
but seem isolated from the others, maybe suggesting physical differences,
e.g. a lower interstellar absorption.
The arc-like shape, identified by region 5
red points, is characteristic of temperature variations
for a plasma in pressure equilibrium \citep{miceli05}.
It seems also present in the off-cloud region 3, though less evident.


\begin{figure}
\epsscale{1.3}
\plotone{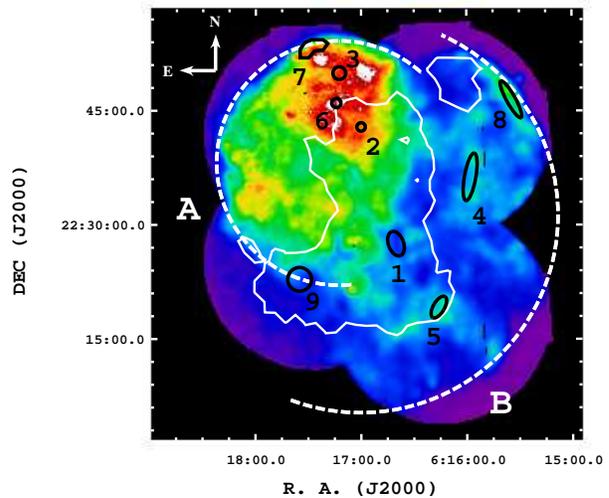}
\caption{X-ray count rate image of IC443 in the 0.3--1.4 keV energy band,
on which we superimposed the nine regions selected
for spectral analysis and the median photon energy contour (white)
corresponding to E$_{50}$=0.97 keV. The dashed semicircles mark
subshell A and subshell B structures. The color bar is the
same as in Fig.~\ref{fig:xima}.}
\label{fig:specreg}
\end{figure}


\begin{figure}
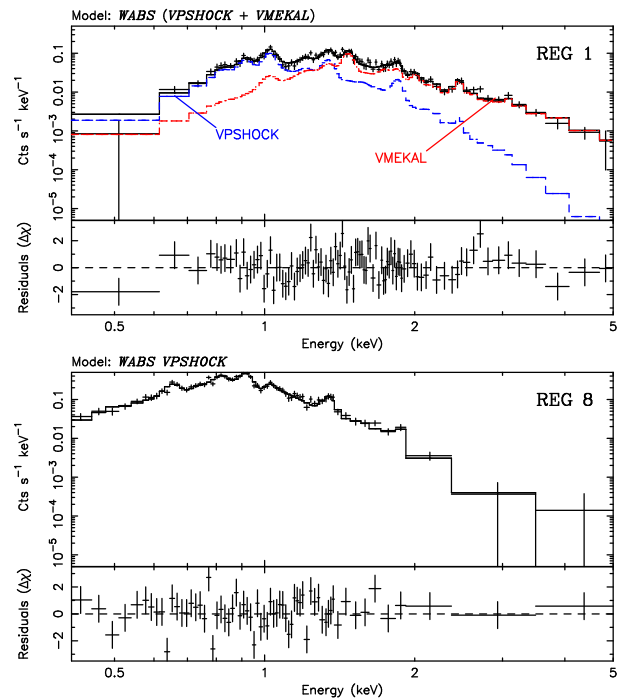

\centering
\includegraphics[angle=270,scale=0.45]{f6a.eps}
\includegraphics[angle=270,scale=0.45]{f6b.eps}
\caption{\xmm\ EPIC MOS spectra representative of IC443 thermal emission. 
The top figure shows the spectrum of reg 1 with the corresponding two components best fit model
and, in the lower panel, the residuals distribution. The blue dashed line represents 
the cold spectral component, while the red dotted line represents the hot one.
The bottom figure shows the spectrum of reg 8 with its corresponding single component best fit model
and, in the lower panel, the residuals distribution.} \label{fig:specs}
\end{figure}


\begin{deluxetable*}{lccccccccc}
\tabletypesize{\scriptsize} \tablecaption{Dimensions, counts and best-fit
parameter values for the nine spectral regions. For each spectrum we also report
the derived filling factor for each spectral component. Quantities in square
brackets are held fixed at the given value. All the reported errors
are at the 90\% confidence level.\label{tab:specres}}
\tablewidth{0pt}
\tablehead{ \colhead{Parameters} & \colhead{Reg~1}
& \colhead{Reg~2} & \colhead{Reg~3}
                       & \colhead{Reg~4} & \colhead{Reg~5} & \colhead{Reg~6}
               & \colhead{Reg~7} & \colhead{Reg~8} & \colhead{Reg~9} }
\startdata
\vspace{0.15cm} Area (arcmin$^2$) $\dotfill$      & 5.2
            & 1.1
            & 2.2
            & 7.5
            & 2.3
            & 1.1
            & 5.5
            & 5.0
            & 7.1 \\
\vspace{0.15cm} Counts (10$^3$) $\dotfill$        &  5.6
            & 6.9
            & 5.9
            & 5.8
            & 6.3
            & 7.0
            & 4.6
            & 4.8
            & 6.7 \\
\vspace{0.15cm} \nh\ (\e{22}\cm{-2}) $\dotfill$      & 0.96$_{-0.2
}^{+0.17}$
                        & 0.74$_{-0.13}^{+0.11}$
            & 0.56$_{-0.10}^{+0.12}$
            & 0.43$_{-0.03}^{+0.06}$
            & 0.45$_{-0.05}^{+0.05}$
            & 0.64$_{-0.06}^{+0.07}$
            & 0.57$_{-0.16}^{+0.12}$
            & 0.58$_{-0.07}^{+0.07}$
            & 0.64$_{-0.13}^{+0.16}$   \\
\cutinhead{Cold component} \vspace{0.15cm} kT$_s$ (keV) $\dotfill$
& 0.37$_{-0.04}^{+0.07}$
                        & 0.58$_{-0.04}^{+0.03}$
            & 0.29$_{-0.03}^{+0.02}$
            & 0.31$_{-0.03}^{+0.06}$
            & 0.64$_{-0.04}^{+0.04}$
            & 0.358$_{-0.012}^{+0.020}$
                        & 0.29$_{-0.06}^{+0.07}$
            & 0.31$_{-0.04}^{+0.07}$
                        & 0.66$_{-0.07}^{+0.06}$   \\
\vspace{0.15cm} $\tau$ (\e{12}\cms) $\dotfill$      &  48 ($\geq$2)
                        &  2.0$_{-1.2}^{+5}$
            &  50 ($\geq$2)
            &  2.3 ($\geq$0.7)
            &  CIE
            &  50 ($\geq$3.3)
            &  0.21$_{-0.10}^{+0.8 }$
            &  1.0$_{-0.7 }^{+9   }$
            &  1.2$_{-0.8}^{+3}$        \\

\vspace{0.15cm} $F_s$\tablenotemark{(a)} $\dotfill$
&4$_{-3}^{+4}$
            &6$\pm3$
            &11$_{-7}^{+20}$
            &8$\pm4$
                &2.3$\pm0.8$
            &7.0$_{-2}^{+1.8}$
            &16$_{-9}^{+30}$
            &70$_{-30}^{+50}$
            &2,1$\pm1,8$ \\
\vspace{0.15cm} O/O$_{\odot}$ $\dotfill$       & [ 1 ]
                        & 4.3$_{-1.6}^{+4}$
                    & 1.3$_{-0.7}^{+3}$
            & 0.44$_{-0.10}^{+0.17}$
            & 2.6$_{-1.0}^{+2}$
            & 0.8$_{-0.3}^{+0.5}$
            & 0.8$_{-0.3}^{+0.5}$
            & 0.52$_{-0.16}^{+0.3}$
            & [ 1 ]  \\
\vspace{0.15cm} Ne/Ne$_{\odot}$ $\dotfill$      & 1.9$_{-1.0}^{+8}$
                        & 3.5$_{-1.2}^{+2}$
            & 1.2$_{-0.6}^{+4}$
            & 0.45$_{-0.11}^{+0.16}$
            & 2.8$_{-0.9}^{+1.9}$
            & 1.0$_{-0.3}^{+0.5}$
            & 0.9$_{-0.2}^{+0.3}$
            & 0.7$_{-0.2}^{+0.4}$
            & 2.6$_{-0.9}^{+6}$  \\
\vspace{0.15cm} Mg/Mg$_{\odot}$ $\dotfill$      & 1.0$_{-0.5}^{+6}$
                        & 1.3$_{-0.3}^{+0.8}$
            & 0.7$_{-0.4}^{+0.2}$
            & 0.32$_{-0.11}^{+0.16}$
            & 1.5$\pm0.5$
            & 0.48$_{-0.12}^{+0.2}$
            & 0.6$_{-0.2}^{+0.3}$
            & 0.50$_{-0.17}^{+0.3}$
            & 1.4$_{-0.5}^{+6}$  \\
\vspace{0.15cm} Si/Si$_{\odot}$ $\dotfill$       & [ 1 ]
                        &  [ 1 ]
            &  [ 1 ]
            &  [ 1 ]
            & 0.4$_{-0.16}^{+0.2}$
            &  [ 1 ]
            & 1.1$_{-0.6}^{+1.4}$
            & 0.5$_{-0.3}^{+0.5}$
            &  [ 1 ]  \\
\vspace{0.15cm} S/S$_{\odot}$ $\dotfill$      &  [ 1 ]
            &  [ 1 ]
            &  [ 1 ]
            &  [ 1 ]
            &  0.13 ($\leq$0.4)
            &  [ 1 ]
            &  [ 1 ]
            &  [ 1 ]
            &  [ 1 ] \\
\vspace{0.15cm} Fe/Fe$_{\odot}$ $\dotfill$      &
0.4$_{-0.2}^{+1.3}$
                        & 0.9$_{-0.4}^{+0.8}$
            & 1.3$_{-0.6}^{+3}$
            & 0.32$_{-0.06}^{+0.09}$
            & 0.15$_{-0.06}^{+0.13}$
            & 0.51$_{-0.14}^{+0.3}$
            & 0.8$\pm0.2$
            & 0.45$_{-0.11}^{+0.2}$
            & 0.5$_{-0.2}^{+1.6}$  \\
\vspace{0.15cm} filling factor $f_s$ $\dotfill$        &   0.15
            & 0.42
            & 0.26
            & 1.0
            & 1.0
            & 0.52
            & 1.0
            & 1.0
            & 0.2 \\	    
\cutinhead{ Hot component} \vspace{0.15cm} kT$_h$ (keV) $\dotfill$
& 1.11$_{-0.16}^{+0.3 }$
                        & 1.78$_{-0.18}^{+0.13}$
            & 1.01$_{-0.06}^{+0.13}$
            &  ---
            &  ---
            & 1.4$_{-0.3 }^{+0.8 }$
            &  ---
            &  ---
            & 1.73$_{-0.13}^{+0.19}$    \\
\vspace{0.15cm} $F_h$\tablenotemark{(a)} $\dotfill$      &
1.4$_{-0.7}^{+0.8}$
            & 0.43$_{-0.16}^{+0.6}$
            & 1.8$_{-1.2}^{+0.8}$
            & ---
            & ---
            & 0.42$_{-0.19}^{+0.3}$
            & ---
            & ---
            & 0.8$_{-0.6}^{+0.5}$ \\
\vspace{0.15cm} Mg/Mg$_{\odot}$ $\dotfill$       & 2.3$_{-1.1}^{+4}$
                        & $>$5
            & 0.8$_{-0.6}^{+0.8}$
            & ---
            & ---
            & $\geq$2
            & ---
            & ---
            & $\geq$4 \\
\vspace{0.15cm} Si/Si$_{\odot}$ $\dotfill$       & 0.5$\pm0.3$
                        & $\geq$3
            & 1.0$\pm0.5$
            & ---
            & ---
            &  $\geq$5
            & ---
            & ---
            & 1.7$_{-0.6}^{+5}$  \\
\vspace{0.15cm} S/S$_{\odot}$ $\dotfill$       & 0.7$_{-0.3}^{+0.7}$
            &  $\geq$3
            &  1.0$\pm0.6$
            &  ---
            &  ---
            &  $\geq$3
            &  ---
            &  ---
            &  1.3$_{-0.5}^{+2}$\\
\vspace{0.15cm} Fe/Fe$_{\odot}$ $\dotfill$      & $\leq$0.06
                        & $\leq$1.3
            & 0.5$_{-0.3}^{+0.4}$
            & ---
            & ---
            & $\leq$1.2
            & ---
            & ---
            & $\leq$0.07  \\
\vspace{0.15cm} filling factor $f_h$ $\dotfill$        &   0.85
            & 0.58
            & 0.74
            & ---
            & ---
            & 0.74
            & ---
            & ---
            & 0.8 \\
$\chi^2$/dof $\dotfill$      &  152/150
                        &  147/125
            &   98/85
            &   76/89
            &  129/96
            &  151/121
            &  103/79
            &  113/99
            &  119/116 \\
\enddata
\tablenotetext{(a)}{Unabsorbed flux in the 0.3--5.0 keV energy band
for the cold ($s$) and the hot ($h$) thermal components. The unit is
\e{-12}\ergs}
\end{deluxetable*}


\subsection{Spectral Analysis}\label{sec:spec}

The \xmm\ data allowed us to significantly improve
the spatial resolution of the spectral analysis
with respect to past works about IC443 \citep{petre88,kawasaki02}.\\
We analyzed spectra extracted from the regions
marked in Fig.~\ref{fig:specreg}. These regions were
selected so as to represent the main
features of the cold plasma,
as obtained from imaging analysis.
We focused on the following aspects:
(a) to verify the presence of a column density gradient across the FOV
  and its correspondence with median photon energy variations (regions 1--5);
(b) to characterize physically the partial shell structure (regions 7 and 8)
  and the site of interaction with the molecular ring (region 9);
(c) to probe the ISM properties in the northeastern quadrant,
  where the surface brightness has its peaks (regions 2, 3, 6 and 7),
  and in the western one, never studied before (regions 4, 5 and 8).\\
Regions shapes and sizes were determined on the basis of statistical
and homogeneity criteria so as to have minimum $\sim$5000 counts
and small fluctuations of the median photon energy ($<$5\%).

Before performing our analysis, we corrected for the vignetting effect
assigning a weight to each event (task \textit{evigweight})
and then we extracted source and background spectra
from the new events list. This procedure changes the spectral
shape of the instrumental background, which is not vignetted,
so it is necessary to extract background spectra
in the same regions as the observed source data.
For fitting we used the sum of MOS1 and MOS2 spectra,
the average of on-axis ARFs,
weighted for the instruments exposure time,
and the MOS1 canned response matrix m1\_110\_im\_pall\_v1.2.rmf.
The spectra were grouped with a minimum of 25 counts per bin and
the $\chi^2$ statistic was used.
We first applied a single thermal component model, testing both for
ionization non-equilibrium (VPSHOCK, Borkowski et al. 2001) and
equilibrium (VMEKAL) cases, and, when needed, we added a second
thermal component to achieve a good fit. We left free to vary O, Ne,
Mg and Fe abundances of the cold component and Mg, Si, S and Fe 
abundances of the hot component. The two components fits with linked 
abundances give worse results than the fits with indipendent abundances.

Figure~\ref{fig:specs} shows two representative spectra, extracted
from regions 1 and 8, with their corresponding best fit models and
residuals. Best fit parameters of all the analyzed regions are
presented in Table~\ref{tab:specres} with the corresponding errors
at the 90\% confidence limit \citep{lampton76}.\\
In the eastern part (regions 1,2,3,6 and 9 in Table~\ref{tab:specres})
IC443 thermal emission in the 0.3--5.0 keV band
can be generally described with a two temperature model: the colder
(kT$\sim$0.3~keV) is characterized by solar abundances
\citep{anders89} and a ionization timescale $\tau\geq$10$^{12}$
cm$^{-3}$ s$^{-1}$, very close to equilibrium conditions, the hotter
(kT$\geq$1.0~keV) by overabundant Mg, Si and S and ion equilibrium.
On the other side, we found the spectra from western regions
(number 4,5 and 8 in Table~\ref{tab:specres}) to be
adequately modeled by a single soft component.

The reduced \chisq\ probability is above 5\% for all regions, except
for region 5 and 9, where we have \chisq$_{\nu}$=1.3 (1.3\%) and
\chisq$_{\nu}$=1.2 (3\%).

\section{Discussion}\label{sec:discuss}
\subsection{Global properties of the X-ray emitting cold plasma}\label{sec:global}

The soft X-ray emission of IC443 shows high values of ionization timescale
and a normal chemical composition,
which suggests an origin from shocked dense circumstellar or ISM material.

Previous works about IC443 had left some ambiguity on the plasma ionization 
state.
The work of \citet{petre88}, based on \textit{Einstein} data,
was not able to discriminate between equilibrium and non equilibrium
ionization scenarios. Afterward, the analysis of ROSAT observations,
made by \citet{asaoka94}, described the X-ray emission with a two temperatures
($\sim$0.3 and $\sim$1.0 keV) Raymond-Smith model, but the data resolution
did not allow to check for the presence of non-equilibrium conditions.
On the basis of ASCA data, \citet{kawasaki02} suggested
a high degree of non-equilibrium  ($\tau$$\sim$\e{11}\cms)
to explain the soft X-ray thermal structure of IC443 in its NE quadrant.\\
The temperature obtained in the selected regions are generally in agreement
with the results of the ``pure IC443" regions of \citet{asaoka94}. Moreover, 
unlike \citet{kawasaki02} our spectral results give a soft X-ray component
near the equilibrium condition ($\tau$$\sim$\e{12}\cms) for all regions
except region 7, in which we have $\tau$$\sim$\e{11}\cms.
The hot component is fully equilibrated.

To infer important values of the gas parameters, such as the mean
electron density and the total mass of the shocked ISM, we assumed a
shell-like morphology and typical cosmic abundances ($n_e$$\sim$1.2
$n_H$) for the soft component. We took into account IC443 double shell
structure, so we approximated the X-ray emitting volume $V$ as two
hemispheres with radii R$_A$=16' ($\sim$7 pc) for the
eastern side (subshell A in Fig.~\ref{fig:specreg}) and R$_B$=26'
($\sim$11 pc) for the western side (subshell B in
Fig.~\ref{fig:specreg}), 
following the results of \citet{braun86} and \citet{cheva99}.

The electron density is thus given by $n_e^2\sim{\rm EM}_s$/ (1.2$f_sV$),
where EM$_s$ and $f_s$ are the emission measure and the soft component
filling factor.\\
For those regions in which we resolved only a soft thermal component,
the standard shell geometry was adopted: we assumed that
the emitting plasma is confined in a thin shell with a thickness of $\sim$1~pc.
To evaluate the filling factors of the other regions
we made the assumption of pressure equilibrium between the soft and the hard component
($n_sT_s$$\sim$$n_hT_h$), deriving
$f_s^{-1}$=1+(EM$_h$/EM$_s$)(T$_h$/T$_s$)$^2$ (see e.g. \citealt{bocchino99}).
The derived filling factor values are listed in Tab.~\ref{tab:specres}.

\begin{figure}
\epsscale{1.2} 
\plotone{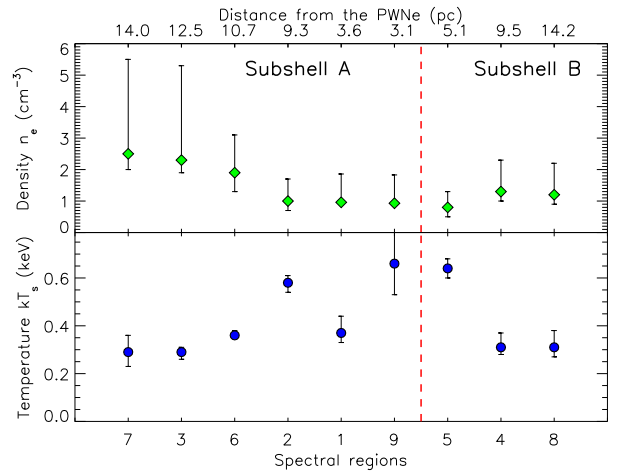} 
\caption{Electron mean density n$_e$ (\emph{top panel}) and
temperature kT$_{\rm s}$ (\emph{bottom panel}) of the cold X-ray emitting plasma
in the nine spectral regions indicated on the X-axis.
The density values were derived from spectral analysis
as the square root of the emission measure divided 
by 1.2 times the volume occupied by the cold plasma.
Subshell A regions (on the left of the vertical dashed line)
and subshell B regions (on the right of the vertical dashed line) 
are plotted in order of distance from the PWNe.}
\label{fig:ne}
\end{figure}

Fig.~\ref{fig:ne} shows electron densities and temperatures for each spectral
region with the corresponding 90\% error bars. Regions are plotted
in order of distance from the PWNe.\\
Subshell B mean density, with a value of $\sim$1~cm$^{-3}$, appears homogeneous.
In subshell A regions nearest to the rim (reg~7,~3,~6) there is an indication
of a density factor two times larger than in the other regions. 
This is consistent with an outwardly increasing density profile near the rim,
which points toward the region of the bright very soft shell and optical filaments.\\
These results can be explained consistently with a shock which is expanding
in a homogeneous and low density environment in the west,
while it has been decelerated by the encounter with a density gradient in the east,
located in a thin region of $\sim$4~pc near the rim.

We made a rough estimation of the total soft X-ray emitting mass
taking  for each subshell the average of the density values,
$<n>_A$ and $<n>_B$, and assuming that the blast wave shell is 2'
thick. We found M$_X$$\sim$30~\msun, which is a small fraction of the predicted
mass of the dense molecular cloud (M$\lesssim$\e{4}~\msun,
\citealt{cornett77}) and the \hi\ cloud (M$\sim$\e{2}-\e{3}~\msun,
\citealt{denoyer78}).
This is reasonable, also in the light of our results about the presence
of a density gradient through most of the IC443 rim (see
sect.~\ref{sec:gradient}), since the material emitting in the
X-rays is just the outskirts of large clouds.


\subsection{Interaction with the environment}\label{sec:gradient}

\begin{figure}
\epsscale{1.2} \plotone{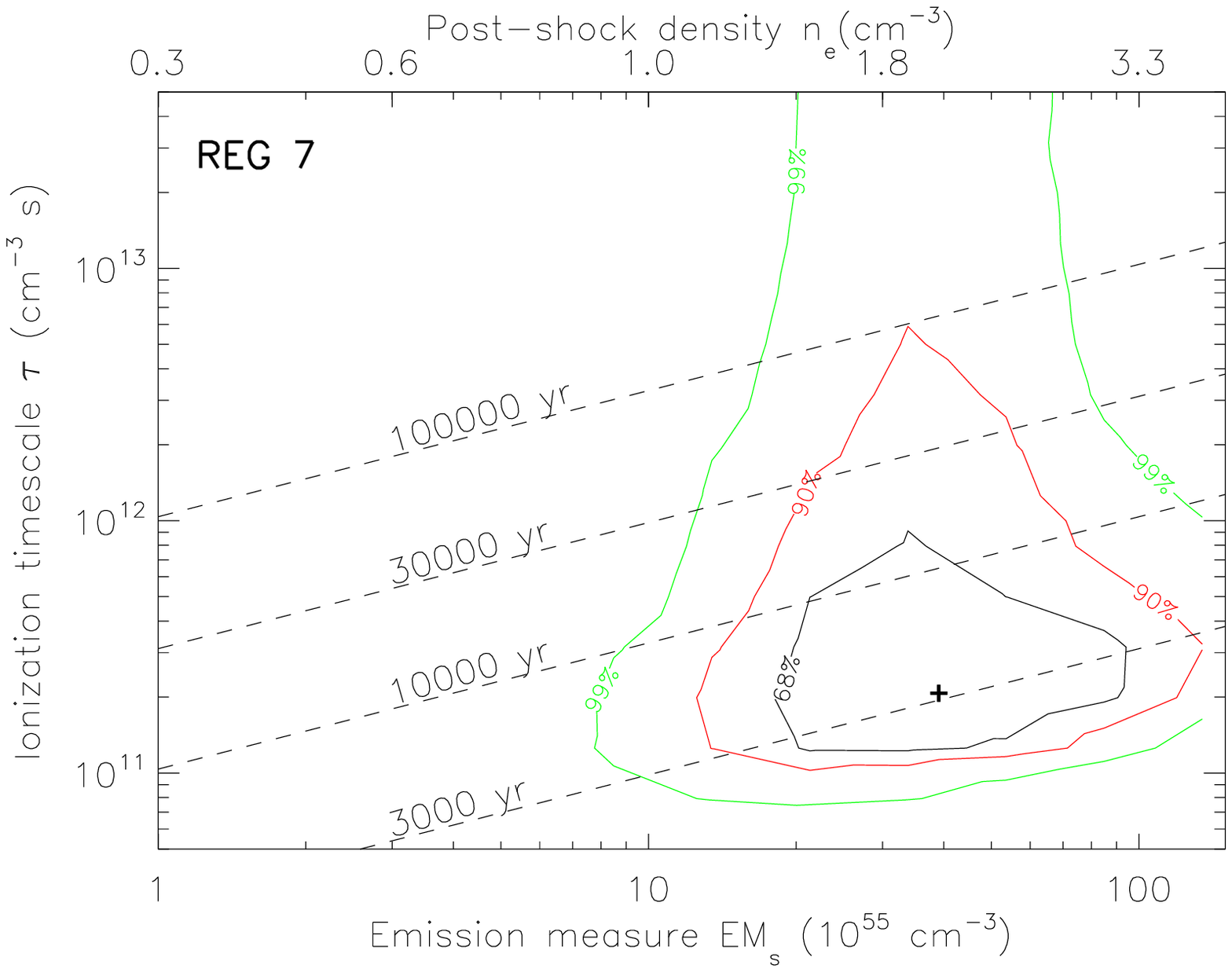} \plotone{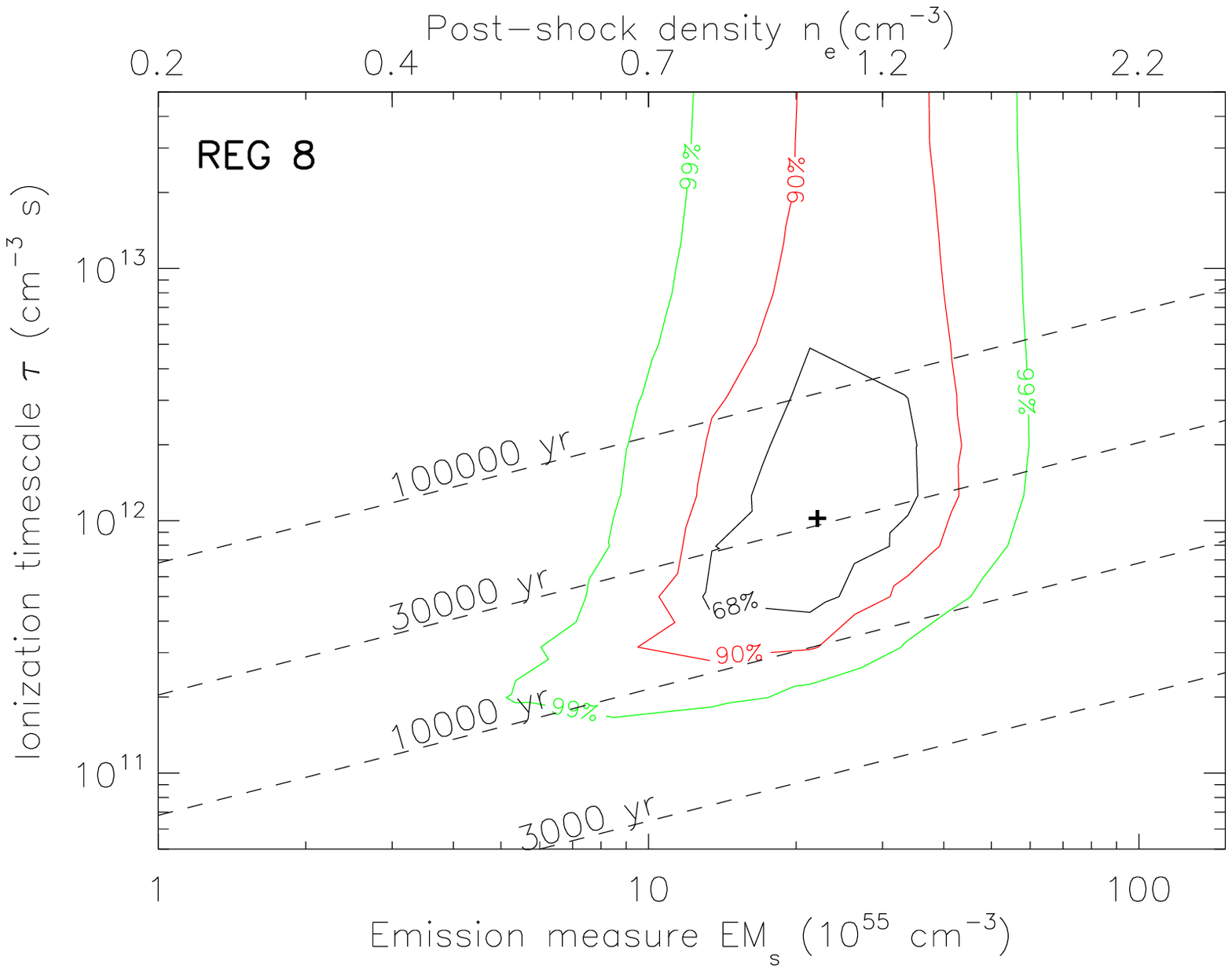} \plotone{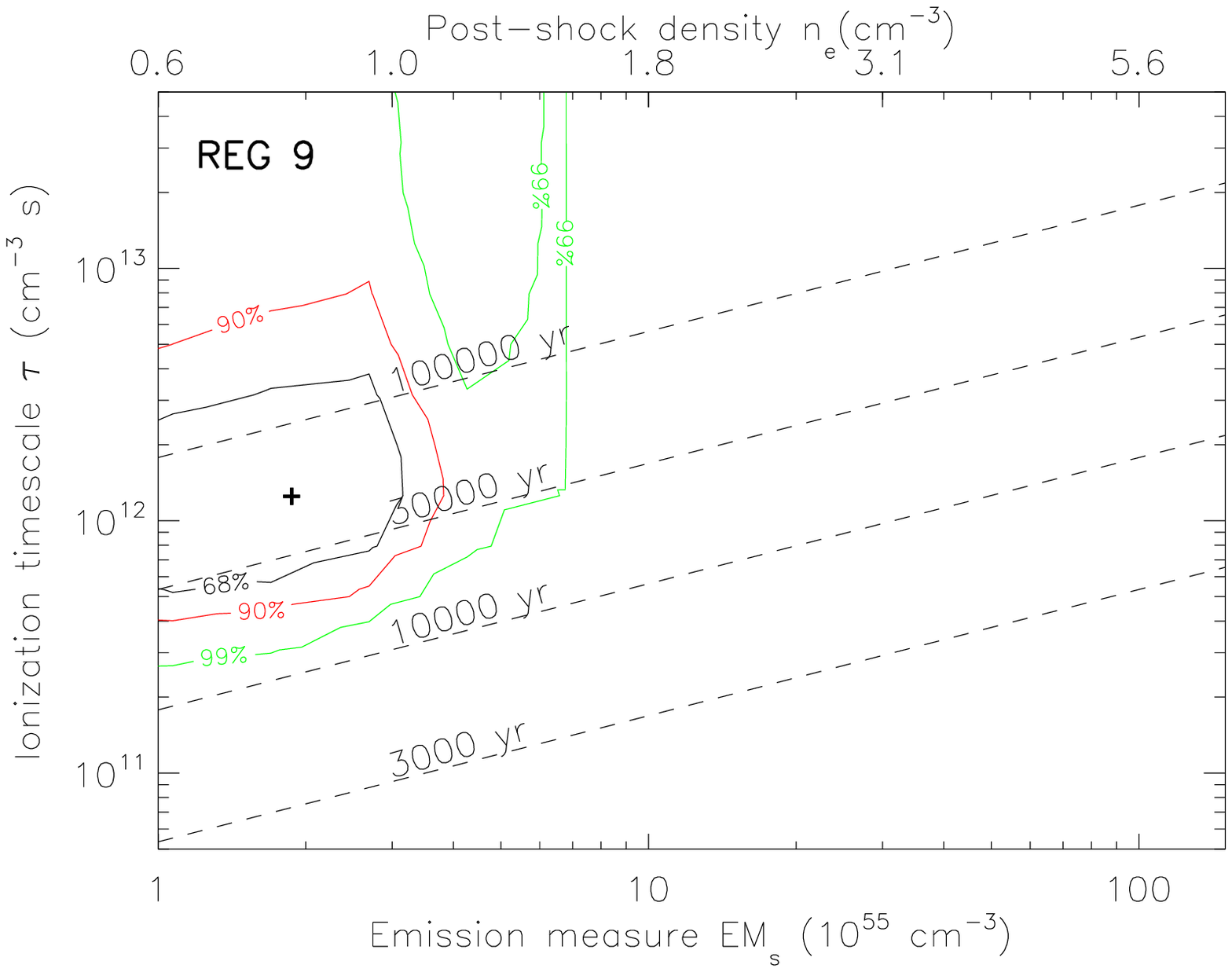}
\caption{Confidence level $\chi^2$ contours (68\%, 90\% and 99\%) in
the EM$_{\rm soft}$-$\tau$ parameter space for three spectral
regions (7, 8 and 9). On the top X-axis are reported the
corresponding post-shock electron densities. We show the range of
ages which are compatible with the values derived from the fit.}
\label{fig:tau_ne}
\end{figure}

We further investigated the evolutionary stage of the X-ray emitting
mass in the region between the remnant and the molecular clouds, in
order to pinpoint possible effects of the interaction with these
large structures. In Fig.~\ref{fig:tau_ne}, we display $\chi^2$
contours in the ionization time $\tau$ vs. emission measure EM$_{s}$
space for three regions located in sites of putative strong interaction
of the remnant with the environment, namely region 7 (interaction
with the neutral clouds), 8 (interaction with ISM) and 9
(interaction with the southern molecular ring). We also show the
post-shock electron density $n_e$ and the age of the emitting plasma
computed using the relation $\tau = n_e t$, where $t$ is the time
since the passage of the shock and $n_e$ is assumed constant after
the shock. The age calculated in this way is a very rough estimate
and it is more useful for comparing different regions than for the
evaluation of the true age.

We stress that for all the regions, the ionization time values are
compatible with collisional ionization equilibrium (CIE) at the
$3\sigma$ level. However, we note that region 7 has a best-fit
ionization time which is $\sim 5$ times lower than 8 and 9 and an
age value which is 10 times lower, thus opening the possibility that
the plasma was more recently hit at NE. This is in agreement with a
location for the SN near the current position of the pulsar wind
nebula (PWNe), which is also supported by the higher values of
kT$_s$ for the regions near the PWNe (e.g. regions 1, 5, 9) than for
the regions far from it (e.g. regions 6, 7, 8). According to this
interpretation, therefore, the regions near the explosion are hotter and
more evolved than the regions far from it, because they were shocked at
higher speed long time ago. From this point of view, however, it is
difficult to explain why the two fronts regions (7 in NE and 8 in NW)
have also different best-fit values for ionization time and age,
since they are probably located at the similar distances from the
explosion site.

From the condition $\rho$v$^2$=constant, we derived an independent
estimate of the evolutionary age of region 7. The shock velocity
corresponding to plasma temperatures of 0.3 keV is v$_{sh}$$\sim$450
km s$^{-1}$. The density of the X-ray emitting plasma that we have
calculated is 2.5 cm$^{-3}$. For the \hi\ cloud we took density and
velocity values from the literature: $n$$\sim$10~cm$^{-3}$ and
v$\sim$100~km~s$^{-1}$ \citep{rho01}. Optical and X-ray brightness
peaks are 23''(2 pc) far apart, so we assumed an exponential density
distribution with a scale height of 0.15 pc. The time elapsed since
the shock front has encountered the cloud is thus $\sim$2\ee{3} yr,
in agreement with the hypothesis of a recent interaction.

Alternatively, the low ionization time of region 7 can be explained by
the fact that the extraction region includes plasma at very different
densities (e.g. in case of strong density gradient). In this case it is
possible that the VPSHOCK models used in the fit try to reproduce the
coldest plasma with a low ionization scale. In fact, we have seen in Fig. \ref{fig:ne} that
the density in this region increases by a factor of 2.5 in 5 pc
(from $n_e\sim$1 in reg~1 to $n_e\sim$2.5 in reg~7), and
that there is an additional compression of a factor of $\sim$4 in 2 pc
between the X-ray and optical filaments, so this interpretation is realistic.

A more accurate estimate of the ionization age in both regions 
7 and 8 would need more data.


\begin{figure*}
\epsscale{0.88} 
\plotone{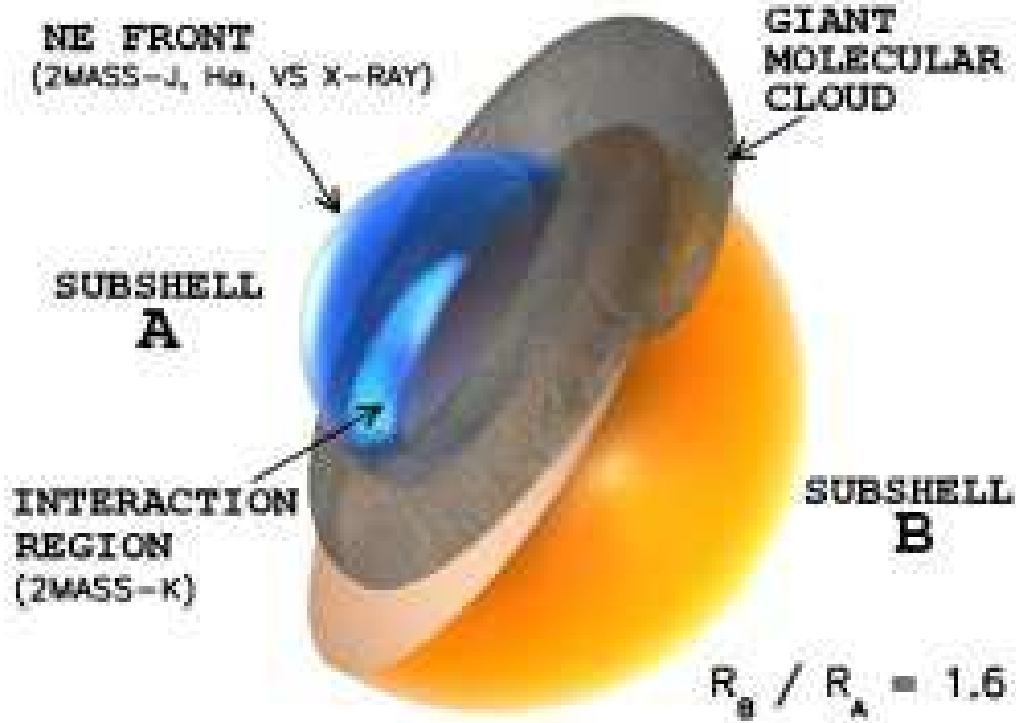}
\caption{Two-shells model describing IC443 morphology. 
The dark ring represents the giant molecular cloud of
\citet{cornett77}.
The blue hemisphere, labelled as subshell A, 
represents the shock front in the eastern region.
In the north-east, where it has been confined by the encounter 
with the neutral \hi\ cloud of \citet{denoyer78},
the shock front is traced by optical, infrared (2MASS-J energy band) 
and very soft X-ray (0.3--0.5~keV energy band) emission. 
In the south-east the light blue strip indicates the region of interaction
with the molecular cloud, which instead is bright in the 2MASS-K energy
band~\citep{rho01}.
The orange hemisphere, labelled as subshell B, represents the shock front in the
western region, where it is expanding in a homogeneous and less dense medium.
The ratio of the shells radii is 1.6, as derived by our X-ray images.} 
\label{fig:3d}
\end{figure*}


\subsection{Absorption from molecular clouds}\label{sec:absorption}

The spatial overlap in the image plane between high median energy values
and molecular emission, shown in Fig.~\ref{fig:midrate},
together with the median/count rate relation suggest an increase
of the foreground absorption in the direction of CO and H$_2$ emission.

\begin{figure}
\centering
\includegraphics[angle=270, scale=0.35]{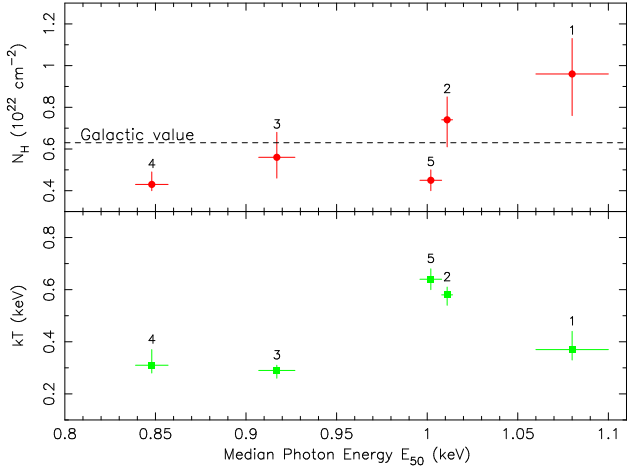}
\caption{Column
density \nh\ (\textit{top panel}) and electron temperature k$_{\rm
B}$T (\textit{bottom panel}) vs. the median photon energy of the
indicated spectral regions.The dashed line in the top panel indicate
the absorption column mean value toward IC443 direction.} 
\label{fig:plmid}
\end{figure}

Using the median photon energy as a spectral indicator, we selected
five representative regions (1, 2, 3, 4 and 5 in Fig.~\ref{fig:specreg})
to study temperature and column density distribution
across the FOV. Our best fit results are plotted in Fig.~\ref{fig:plmid}
with the corresponding error bars at the 90\% confidence level.\\
They confirm that higher \nh\ values correspond to higher median energies and identify
an increasing trend of the column density. Moreover, the best fit absorbing column
of region 1 is significantly higher than the mean Galactic absorption toward this
portion of sky (N$_{\rm H}^{\rm Gal}\sim$ 6.3\ee{21}\cm{-2}, \citealt{dickey90}),
indicating the presence of additional absorbing material.
We estimate that it contributes an equivalent column density of $\Delta$\nh$\sim$5\ee{21}\cm{-2}.
The spatial distribution of the obscuring material and the corresponding \nh\ values
point to the identification with the unperturbed cloud of \citet{cornett77}.
Our X-ray data strongly suggest that it is located in front of IC443,
between the remnant and the observer.

The southern sinuous ridge shows the same E$_{50}$ values of the
quiescent cloud, indicating similar hydrogen column density. In the
median energy map of Fig.~\ref{fig:q50}, the southern ridge and the
quiescent cloud appear as a single complex. We argue that both
probably belong to the same cloud structure, which has been reached
by the IC443 shock in several location of the south side
(corresponding to clump A to G of \citealt{dickman92}), while it is
mostly unshocked elsewhere, in particular in the on-cloud regions
(Fig.~\ref{fig:q50}) where there is no maser or H$_2$ infrared
emission. \citet{dickman92} suggested that the shocked material is
in a $\sim$9$\times$7~pc ring tilted by $\sim$50$^o$ from the line
of sight and that it is expanding, presumably as a result of the
interaction. We expand this picture by arguing that the Dickman ring
represents only the shocked part of a large molecular cloud complex
shaped as a large tilted torus, which is obscuring the background
SNR X-ray emission.

In Fig.~\ref{fig:3d} we show the proposed model for the environments of IC443.
This model is supported by the comparison (see
Fig. \ref{fig:q50}) between our median energy image and the contour
of \citet{burton88}, which traces the boundaries of the shocked gas,
and the CO contours indicating quiescent dense material. While the
southern and western on-cloud regions are characterized by high
median energy values (E$_{50}$$>$0.98 keV), which locate the
molecular ring in the foreground, the northeastern regions have
E$_{50}$$\sim$0.9 keV, indicating lower extinction and thus,
suggesting that the large torus lies in the background, as
expected in the case of a tilted geometry encompassing the whole
SNR. This picture is in agreement with kinematic velocities measured
in clumps A to H by \citet{dickman92}, and by the more recent study
of clump H and H$_2$ infrared emission of \citet{rho01}. In
particular, the latter authors detect weak 2MASS-K filaments in a
region called ``NE front'', which includes clump H and is separated
from the bright optical and radio NE rim. They argue that the clump
H/NE front region is part of the molecular cloud interaction, as the
sinuous southern ridge, and the fact that it is an ``off-cloud''
region according to our Fig.~\ref{fig:plmid} suggests that it
represents the interaction of the SNR shock with the part of the
torus lying behind the SNR.

The comparison between the 0.3--0.5~keV map (Fig.~\ref{fig:xima2})
and the median energy map (Fig.~\ref{fig:q50}) shows that the very soft shell is not
correlated with absorption in the east of the remnant, thus indicating that this shell is
not entirely an effect of absorption.

The median energy map allowed us to investigate the issue of the
location of the nearby SNR G189.6+3.3.  This remnant was located by
\citet{asaoka94} in front of IC443, causing a dark lane of
absorption equivalent to $\Delta$\nh$\sim$2\ee{21}\cm{-2}.  This
would raise the value of the median energy of
$\Delta$E$_{50}$$\sim$80~eV.  Our results does not support strong
absorption from the shell of G189.6+3.3.  We conclude that either its
contribution to \nh\ should be small compared to that of the CO
cloud or the SNR could be in the background.

Finally, from the spectral results presented in Table~\ref{tab:specres}, we argue that the
peculiar location of region 5 in the  temperature vs.~median energy
plot (Fig.~\ref{fig:plmid}, {\it lower panel}) is an effect of high
temperature rather than column density.  This is expected, since for \nh$<$
5\ee{21}\cm{-2} our median energy map is more sensitive to temperature
variations than for higher \nh values.


\section{Conclusions}\label{sec:end}
We presented results from \xmm\ observations of the Galactic SNR IC443.
High resolution image in the 0.3-0.5 keV energy band unveils 
a partial limb-brightened X-ray shell, which resembles on small angular scales
the optical and radio morphology in the NE quadrant and which is
different from the shell structure previously reported by \citet{kawasaki02}.
We concluded that it spatially traces the
interaction of the shock with a density gradient,
which is part of the neutral northeastern cloud studied by \citet{denoyer78}.

The spatially resolved spectral analysis of IC443 measured large
column density variations across the remnant, and the median photon
energy distribution shows that they are spatially coincident with
molecular emission from dense molecular clouds in the FOV. Our
results confirmed that X-ray emission is strongly absorbed by the
molecular cloud of \citet{cornett77} and by the southern sinuous
ridge of \citet{burton88}.
We did not find evidence of absorption from the SNR G189.6+3.3.\\
On the basis of our median energy map, we proposed that the
southern ring and the more extended molecular cloud are
part of the same structure and that it is
tilted respect to the line of sight, as suggested by \citet{dickman92}.

We performed spectral analysis in 9 bright regions with at least 5000 counts
and a 5\% maximum variation of the median photon energy E$_{50}$.
The X-ray emission from our selected regions
is well described with CIE or near ionization equilibrium models.
Our measured X-ray temperatures and ionization timescales are compatible 
with a density gradient in the NE rim and suggest that the site of the 
SN explosion could be near the PWNe.


\acknowledgements{}
The authors thank the referee for his/her comments and suggestions.
We wish to thank Valerio Guarneri for his collaboration 
to realize Fig.~\ref{fig:3d}.
The CO data were obtained by Mark Heyer and Min Yun and kindly
provided by J.~J.~Lee.

This work was supported by Ministero Istruzione Università e Ricerca (MIUR), 
Agenzia Spaziale Italiana (ASI) and Istituto Nazionale di Astrofisica (INAF).


\bibliographystyle{aa}
\bibliography{myreferences}

\end{document}